# Coarse-Grain Performance Estimator for Heterogeneous Parallel Computing Architectures like Zynq All-Programmable SoC

Daniel Jiménez-González*[†], Carlos Álvarez*[†], Antonio Filgueras[†], Xavier Martorell*[†],
Jan Langer[‡], Juanjo Noguera[‡] and Kees Vissers[‡]
*Universitat Politecnica de Catalunya
[†]Barcelona Supercomputing Center
Email: {daniel.jimenez,carlos.alvarez,antonio.filgueras,xavier.martorell}@bsc.es
[‡]Xilinx USA
Email: {jan.langer,juanjo.noguera,kees.vissers}@xilinx.com

*Abstract*—Heterogeneous computing is emerging as a mandatory requirement for power-efficient system design. With this aim, modern heterogeneous platforms like Zynq All-Programmable SoC, that integrates ARM-based SMP and programmable logic, have been designed. However, those platforms introduce large design cycles consisting on hardware/software partitioning, decisions on granularity and number of hardware accelerators, hardware/software integration, bitstream generation, etc.

This paper presents a performance parallel heterogeneous estimation for systems where hardware/software co-design and run-time heterogeneous task scheduling are key. The results show that the programmer can quickly decide, based only on her/his OmpSs (OpenMP + extensions) application, which is the co-design that achieves nearly optimal heterogeneous parallel performance, based on the methodology presented and considering only synthesis estimation results. The methodology presented reduces the programmer co-design decision from hours to minutes and shows high potential on hardware/software heterogeneous parallel performance estimation on the Zynq All-Programmable SoC.

## I. INTRODUCTION

With the end of Dennard Scaling [7] computer architecture has entered a new era. One main thread followed by several architectures has been to evolve into multi- and many-core systems composed of several identical cores. Another important trend has been the incorporation of large, specialized accelerator systems (mainly evolved from the graphics ecosystem) that efficiently execute single instruction mulitple thread codes.

However, the struggle to squeeze some performance out of a continuously growing number of transistors per chip has, somehow, avoided its most obvious and promising path: the creation of a large number of very specialized accelerators that can, on one side, be really energy efficient and, on the other, make the work faster by eliminating software overheads. Indeed, ASICs, so common only a few years ago, are being progressively discarded in favor of cheaper, more general components that have the essential advantage of short time-to-market cycles. In this sense, new hybrid CPU-FPGA systems can be seen as the future of heterogeneous computing. While being roughly one order of magnitude slower that its ASIC equivalent, FPGAs can be reprogrammed on the fly, and adapt to changing environments. Furthermore, being tightly integrated with general cores, those systems can retain the programmability of common CPUs and join it with the tremendous boost in performance and efficiency that characterizes specialized hardware.

However, in order to be broadly used in the mass-market, those systems still face two important challenges: first, a software ecosystem that facilitates their programmability without burdening the programmer with all the cumbersome details (data transfers, synchronization, memory coherence...) of heterogeneous systems and, second, an easy and fast way to perform a quick decision of the best mapping of all the application components to the most adequate hardware to compute them in a parallel heterogeneous system.

Several works have addressed the first problem and it can be said that parallel programming models (like OmpSs [9] and OpenMP 4.0) can be used to solve it. However the second problem is still a barren field. Indeed, assuming that the first step of selecting which application kernels can be computed by the reconfigurable hardware is done by a programmer, and generating the proper HDL code for them could be done automatically, still remains the problem that a FPGA bitstream generation can take several hours. After that, the whole application should be analyzed only to find out if the hardware-software partition, or the resources distribution between the kernels in the FPGA, are adequate or not for the scheduling policy of the programming model and the heterogeneous system at hand. Any mistake in the selection or any bad guess by the expert programmer means repeating the whole process leading to a trial and error process composed of several hours steps..., in addition to a unexpected parallel heterogeneous performance.

In this paper we address those issues by presenting a way to speed-up the process and help introduce heterogeneous systems into everyday computation. The suggested workflow is designed to estimate the performance of OmpSs applications on any heterogeneous parallel architecture using the execution time information (estimated or not) of the OmpSs tasks running on the processing components of the target architecture. Those heterogeneous architectures are currently target by cyber-physical computing platforms [2], [5] that may combine cluster of nodes with SMP, FPGAs and GPUs. Here, we focus on the heterogeneous parallel performance estimation for the Zynq All-Programmable SoC architecture, that combines ARM-based SMP and a FPGA, and that also includes GPUs in the next generation Zynq UltraScale+ MPSoC[20]. This workflow integrates a simulator which implements the runtime of the OmpSs programming model, and a shared memory coarse-grain component (ARM cores and FPGA accelerators) architecture with local memory (BRAM) for the FPGA accelerators. This simulation is fed by the reports obtained from the high level synthesis tool for the timing information of the hardware accelerators considered by the programmer and a task-based trace generated by a sequential execution of the OmpSs application. The framework simulates the execution of the trace tasks in a data-flow manner as the software runtime of OmpSs would do, considering all the components of the heterogeneous architecture. With this, the simulator can obtain the estimated heterogeneous parallel performance for the application kernels (tasks) in the target system. Finally, once the best alternative is selected based on the simulator results, the bitstream can be effectively generated and executed in order to check the correctness of the conclusions. This process lasts just a few minutes or even seconds to achieve similar results than having to generate the bitstream for each possible mapping and run the application in the real system.

Our methodology currently considers that the programmer is an parallel programmer that only needs to explore few hardware/software codesigns, otherwise a design space exploration strategy should be analyzed to reduce the amount of possible solutions, like using back annotations [11], [19].

So, the contributions of the paper are as follows:

- Light coarse-grain estimation that helps the programmer to have a fast order-of-magnitude decision of the best hardware/software co-design on a heterogeneous parallel system with FPGA devices, reducing the number of bitstreams to be generated.
- *Heterogeneous parallel performance* estimator based on a task-based trace driven simulator that integrates the runtime of the OmpSs programming model and a shared memory coarse-grain component architecture, with local memory for the hardware accelerators.
- A complete framework that avoids the need of placing and routing each FPGA accelerator required by the programmer annotated tasks with target FPGA.

The rest of this paper is organized as follows: Section II presents the related work of the paper. After that section III presents the methodology suggested, section IV its implementation and section V presents the experimental setup. Section VI presents the results obtained. Finally, section VII concludes the article.

## II. RELATED WORK

The work presented in this paper addresses the problem of how to use efficiently, at run-time, the hardware resources of an heterogeneous system selecting the best among the large amount of implementation possibilities that such a system offers. To do so, it relies in the existence of a programming model that allows the integrated programming of heterogeneous systems, predicting the performance of the applications on those systems. Meanwhile there are several programming models dealing with FPGA-based heterogeneous systems [15], [6], [1], [10], [16], [8], there has been less work in the literature regarding the performance prediction of heterogeneous applications. Indeed, most of those works cover the prediction of a kernel performance on an FPGA [13], [14], [18] and only few of them analyze full application performance prediction. The RC amenability test (RAT) [12] is a system that tries to predict the suitability of an application to be ported to an FPGA, in order to avoid the work if the outcome is foreseeable as non successful. Another performance prediction technique [17] addresses the modelling of shared heterogeneous workstations containing reconfigurable computing devices. This methodology chiefly concerns the modelling of system level, multi-FPGA architectures. However, it does not take into account the selection among several different application kernels or the interactions between them in a given parallel application.

Other works propose electronic system level timing and power estimation that combines system-level timing and power estimation techniques with platform-based rapid prototyping [11], [19]. However, the annotated task has to be specified in a particular language and/or has to be mapped to a specific component of the system. In our work, the same task can be annotated in C/C++ language with OpenMP-like task directives, which is the standard shared memory programming model. In addition, tasks can be annotated to be mapped, at run-time, to different components of the heterogeneous system depending on the scheduling policy.

To the best of our knowledge the work presented in this paper is the only one that deals with those kernel selection and performance prediction challenges that require a *run-time* analysis of the application and the prediction of complex, irregular task dependency execution patterns.

## III. METHODOLOGY

In this section we present the methodology proposed in this paper to reduce the developer effort to map complex applications to heterogeneous parallel systems with FPGAs. OmpSs programming model makes easy the programmability of applications with kernels (pieces of code, functions or not) that may be executed in both CPUs and FPGAs [10] as offloaded tasks, transparently to the programmer, by writing simple pragma directives. Note that, as commented above, the automatic generation of the different granularities and the architecture configurations (number of accelerators for each kernel) is beyond of the scope of this paper contribution. Therefore, the granularity of the tasks and target devices where those tasks could be run should be indicated by the programmer. The decision of where those tasks are executed is automatically done at run-time.

Figure 1 shows an example of an OmpSs blocking matrix multiplication. As it can be observed in the first line of the code, the `mxmBlock` kernel has been annotated as it can be executed in both FPGA and SMP. The second line of the code specifies that this function will become a OmpSs task (with input and output dependences) any time it is called in the code. With these annotations, the OmpSs runtime can take care of scheduling different instances of the kernel, when their dependences are ready, in both resources based on availability.

```c
#pragma omp target device(fpga,smp)
#pragma omp task in([BS*BS]A,[BS*BS]B)\
                inout([BS*BS]C)
void mxmBlock( REAL *A, REAL *B, REAL *C)
{
  int i, j, k;
  for (i=0; i < BS; i++)
    for (k=0; k < BS; k++) {
      REAL tmp = A[i*BS+k];
      for (j=0; j < BS; j++)
        C[i*BS+j] += tmp * B[k*BS+j];
    }
}

void matmul(REAL **AA,REAL **BB,REAL **CC,int NB)
{
  int i, j, k;
  for (k = 0; k < NB; k++)
    for(i = 0; i < NB; i++)
      for (j = 0; j < NB; j++)
        mxmBlock(AA[i*NB+k],BB[k*NB+j],\
                CC[i*NB+j]);
}
```

Fig. 1. Matrix multiplication annotated with OmpSs directives. `matmul` is the blocking matrix multiplication function, and `mxmBlock` performs the matrix multiplication of a block.

However, even if the translation from C to HDL is done automatically with Vivado and OmpSs is used to schedule the `mxmBlock` tasks to the best available computing unit, the problem of how to partition the work remains. How many instances of the `mxmBlock` should be implemented in the available hardware? How big should be any of the instances? Is it worth to implement two instances of a different size? Indeed, in the presence of several possible kernels to be mapped to the FPGA when all of them will not fit, which ones have to be mapped to maximize the application performance? The expert programmer may have an idea of which is the best combination and reduce the number of possible implementations to few of them (tens). However, those few implementations may mean hundreds of hours if each of them implies one or more bitstream generations.

In order to answer those questions a coarse-grain performance estimator toolchain has been developed. This performance estimator toolchain combines instrumentation based on source to source compilation, high level synthesis from C code and a heterogeneous task-based dataflow parallel simulator developed to estimate the heterogeneous parallel performance of OmpSs code with heterogeneous tasks. This simulator has Extrae [3] instrumentation support that allows it to generate Paraver [4] traces, allowing the programmers to have an approximate visualization of what one would expect in a real task execution on an heterogeneous system. Extrae is a instrumentation library that generates time events, thread states and communications in a raw trace that can be translated to a Paraver trace format. In this work we have integrated the simulation with a modified version of the Extrae so that Extrae can take the timing of the simulation. Paraver is a tool that allows performance analysis of parallel execution traces at different levels of granularity: thread, task, MPI process, etc.

Figure 2 shows the overall steps of the developed toolchain. The parallel programmer has to provide the OmpSs code with the annotation of the tasks and the granularity he/she wants to evaluate. As we mentioned above, the parallel programmer should have an idea of which are the most potential combinations of tasks, reducing the amount of possible task mappings and granularities. The automatic generation of the different granularities is beyond of the scope of this paper contribution. However, a starting programmer may need to analyze a large number of granularities and mappings, and in this case, a sytem to automatize the design space exploration would be helpful [11].

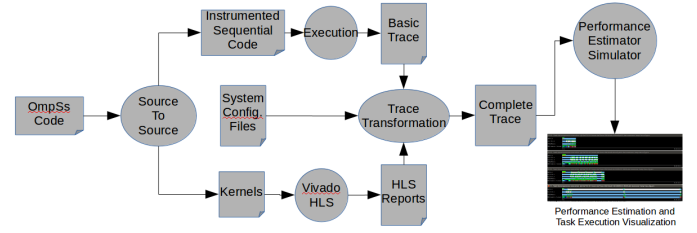

Fig. 2. Coarse-Grain Performance Estimator Toolchain

The first step of the toolchain is to (1) transform the OmpSs code to a sequential instrumented code, and (2) extract the kernel code of each task annotated by the programmer. Both these steps are automatically performed by a source to source compiler from the original OmpSs code. Once the instrumented sequential code has been obtained it is executed in order to obtain a trace of tasks that will be used for the performance estimation. The information contained in the trace will be joined with hardware timing information (estimated cycles and clock frequency) obtained from passing the extracted kernel codes through Vivado HLS and fed to our heterogeneous performance estimator that simulates the dynamic behavior of a preconfigured system (a particular implementation of the application in the Zynq board in our case) and returns not only the estimated time used by the given application in the selected hardware configuration but also a Paraver trace that can be visualized in order to further analyze the possible bottlenecks of the design. The whole cycle only takes few minutes and can be repeated as many times as necessary until all the possibilities have been explored. Finally, the best implementation can be chosen and the time consuming process of the hardware bitstream generation is done only once. The next section further explains how the performance estimation is done accurately enough to obtain useful results for the programmer's hardware/software co-design decision.

## IV. IMPLEMENTATION

In order to obtain enough information from the OmpSs code, this is first transformed into an instrumented sequential code by source to source compilation. During this transformation, the directives of OmpSs are replaced with instrumentation of the tasks to be able to generate a task execution trace that will contain the following basic information: task number, creation time and elapsed execution time in cycles in the CPU based machine, number of dependences of the task, and for each dependence: the data dependence memory address and a label indicating the direction (input, output or inout) of the dependence, and finally, task name for later identification in the performance estimator toolchain.

The basic trace, generated by the execution of the instrumented sequential code, should be completed with further information. First of all, the cost of creation of a task has to be added. Each OmpSs task has a creation cost that is not generated by the instrumented sequential code. Therefore, each task instance of the task execution trace needs to be preceded by its creation cost (creation cost task), that will be run (in the simulation) only in the SMP device (independently if the task is executed in the FPGA or in the SMP). The original task instance in the trace will depend on the new creation cost task. Next, the information of the devices where each task can be executed and the latency of those should be also added. With this objective, the extracted kernels are used in order to obtain the latency of the hardware accelerators of those tasks that can be run, based on programmer annotation, in the FPGA. The latencies estimated for the computation and the input and output transfers are obtained by passing the extracted task code through the Vivado HLS, which, in few seconds, can generate the HDL code and a report with all the information required for this task code:

- Estimated number of cycles of the computation of the task in the FPGA

- Estimated number of cycles spent transferring the input/output parameters of the task to the FPGA

Using that information, each of the task instances that appears in the basic trace is completed with more information that states that the task can be also run in a hardware accelerator, and with the latency of the associated hardware accelerator.

Further specific information, related to the system where the programmer wants to execute the OmpSs code, should be taken into account to complete the trace. For instance, it should be evaluated if the system can overlap input and/or output DMA memory transfers (between the shared memory and the local memory in the accelerators) among different hardware accelerators. However, this analysis only needs to be done once. If the transfers can be done in parallel the input and output data transfer latencies can be added to the computational latency of the hardware accelerator associated to this task. Otherwise, DMA memory transfer tasks will be created and run in a shared hardware resource device to avoid possible overlapping of input/output transfers. Those extra tasks will have dependences with the corresponding tasks run in the device. In the case of our target architecture, the Zynq 706 board, and the current environment analyzed, the input parameters seem to scale with the number of accelerators, but not the output parameters. Figure 3 shows the speedup obtained when using 2 accelerators compared to 1 accelerator to transfer the same amount of input and output memory data: 512Kbytes and 1024Kbytes. Therefore, the time associated with a task running in a hardware accelerator device can be seen as the time of the input data DMA transfer plus the computation time. This information, together with the time this task lasts in a SMP core, will be part of the information of this task in the trace. However, the output DMA memory transfer cost will be represented by an new transfer task that will be run in a shared hardware resource device to avoid possible overlapping of output transfers, since no overlapping seems to be allowed. This output transfer will have an dependence with the original tasks run in the device.

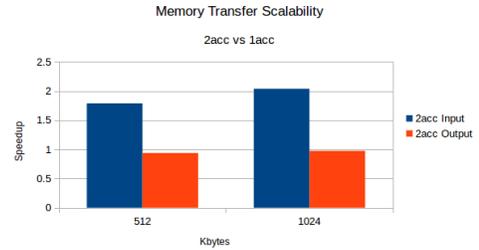

Fig. 3. Speedup of using 2 accelerators vs 1 accelerator for the input/output data transfers on the Zynq 706 Board for two different amounts of data.

On the other hand, each of those DMA transfers has to be programmed in software from the SMP device. This software cost may not be able to be done in parallel since they have to use shared resources. Then, DMA programming tasks (submit tasks) that will be run in a special device, shared among all the hardware accelerators, are created for each input/output transfer. The original task will depend on the input submit tasks and the output submit tasks will depend on the original task.

Once the trace has been completed with all the above information, the heterogeneous parallel architecture performance estimator can simulate the execution of all the tasks (original, creation cost, and DMA related tasks) in a dataflow manner for a given configuration of the hardware. That is, it will take care about the task input, output, and inout dependences and will run them as soon as their dependences are ready and a device that can execute them is available. The task dependency management is done the same way that the OmpSs runtime software system does. The performance estimator toolchain can be run, for an specific transformed trace, under different hardware configurations based on the programmer annotations. As a result, the Paraver traces generated by the estimator will allow the programmer to choose the best estimated combinations of software and hardware accelerators.

## V. EXPERIMENTAL SETUP

Results in Section VI have been obtained on a Zynq All-Programmable SoC 706 board. Timing of the applications has been obtained by instrumenting with `gettimeofday` the part of the code that calls several times the kernel code. Results show the average elapsed execution time of 10 application executions on the Zynq 706 board under linux.

The OmpSs implementation is based on Mercurium 1.99.4 and Nanos++ 0.8. For the hardware compilation branch we have used the Xilinx ISE Design 14.7 and the Vivado HLS 2013.2 tools. The estimator has been developed with support for Extrae Library 2.5.1 and Paraver 4.3.5. The Paraver was used to analyze the estimated execution traces. All OmpSs codes have been compiled with the `arm-xilinx-linux-gnueabi-g++ (Sourcery CodeBench Lite 2011.09-50) 4.6.1` and `arm-xilinx-linux-gnueabi-gcc (Sourcery CodeBench Lite 2011.09-50) 4.6.1` compilers, with `"-O3"` optimization flag.

We show real execution and estimator results for 2 tiled applications: matrix multiply (Figure 1) and cholesky (Figure 4), using different `fpga` task granularities for the tiles (blocks): $64 \times 64$-block single-precision floating point matrix multiply (fine-grained tasks), $128 \times 128$-block single-precision floating point matrix multiply and $64 \times 64$-block double-precision floating point cholesky decomposition. In the case of the cholesky decomposition three out of four of the kernels are annotated to be able to be run in the SMP and also the FPGA. The fourth one has not been considered to be mapped to the FPGA by the programmer. All real data generated by the Vivado HLS has been synthesized with IEEE-754 standard compliance.

```
#pragma omp target device(fpga,smp)
#pragma omp task in([BS*BS]A) inout([BS*BS]C)
void dsyrk(double *A, double *C, int BS);

#pragma omp task inout([BS*BS]A)
void dpotrf( double *A, int t, int BS );

#pragma omp target device(fpga,smp)
#pragma omp task in([BS*BS]A) inout ([BS*BS]B)
void dtrsm( double *A, double *B, int t, int BS);

#pragma omp target device(fpga,smp)
#pragma omp task in([BS*BS]A, [BS*BS]B)\
                inout ([BS*BS]C)
void dtrsm( double *A, double *B, double *C,\
            int t, int BS);

void chol_ll(double **AA, int t, int NB, int BS)
{
  for (int k = 0; k < NB; k++ ) {
    for (int j=0; j<k; j++)
      dsyrk(AA[j*NB+k], AA[k*NB+k], BS);

    dpotrf(AA[k*NB+k], t, BS );

    for (int i = k+1; i < NB; i++)
      for (int j=0; j<k; j++)
        dgemm(AA[j*NB+i],AA[j*NB+k],\
              AA[k*NB+i],t,BS);

    for (int i = k+1; i < NB; i++)
      dtrsm(AA[k*NB+k], AA[k*NB+i], t, BS );
  }
}
```

Fig. 4. Cholesky application annotated with OmpSs directives. Each of the function calls will be a task instance (dsyrk, dtrsm, dtrsm: SMP and FPGA, dpotrf: SMP only).

## VI. RESULTS

In this section, a coarse-grain comparison of the estimator and real execution results is shown. The comparison is done varying relevant aspects in the design of heterogeneous parallel applications for FPGA based architectures, highlighting the analysis time required in our proposed methodology.

In the case of the tiled matrix multiplication we show a performance estimation that evaluates three different possible design decisions. The first one is to select between two different task granularities (64x64 blocks and 128x128 blocks) for the task kernel mxmBlock in Figure 1. Second, it is evaluated the difference between using one or two accelerators for running 64x64 mxmBlock tasks. Having two accelerators for the 128x128-block case has not been considered in the evaluation because the hardware resource estimation for two 128x128-block mxmBlock accelerators indicates that it is not feasible to map them into the programmable logic. Finally, we have considered the performance impact of allowing heterogeneous execution (mxmBlock task is specified with SMP and FPGA) or not.

Figure 5 shows the performance results of the commented cases for both the estimator and the real execution. Results are normalized with respect to the slowest case (one accelerator of 128x128 blocks and with heterogeneous execution - label 1acc 128 + smp in the figure). Although estimator and real execution have different absolute speedups (our estimator does not consider memory hierarchy aspects like cache coherence and pinning of memory pages, neither memory contention, etc.), results show the same speedup trends. That allows the programmer to adapt her/his OmpSs program to have 128x128 mxmBlock tasks with the FPGA as the only target device. This decision can be taken after less than 5 minutes of work (coffee break), that is what the analysis requires under the proposed methodology. Figure 6 shows the time (seconds) in logarithmic scale for the analysis of the configurations under our proposed methodology (left) and the traditional hardware-software design cycle (right). In particular, for the traditional design cycle, we only count the hardware generation of the different accelerators and combinations. The hardware generation time required for the full-analysis is more than 10 hours. On the other hand, the performance estimator toolchain lasts for less than 5 minutes and automatically provides the best choice among the considered configurations.

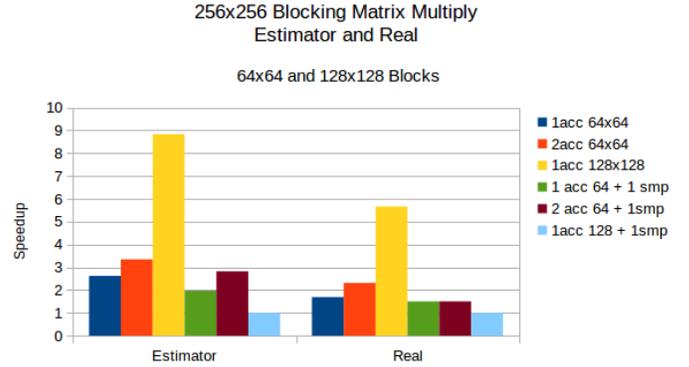

Fig. 5. Estimation and real matrix multiply performance comparison for different hardware configurations of the system and task configurations.

In addition to this decision, the programmer may want to do a depth analysis of the performance estimation using the Paraver traces generated in the estimation process. Paraver traces can be visualized and compared to detect potential bottlenecks in the parallel and heterogeneous execution of the tasks. Figure 7 shows the Paraver view of four estimated task execution traces for four different configurations shown in Figure 5 with the same time scale; from top to bottom: 1 acc 128x128, 2 acc 64x64, 2 acc 64x64 + SMP and 1 acc 128x128 + SMP. Paraver traces show the execution of the tasks (original and additional ones) in the devices, along the time (x-axis). Each Paraver trace shows an horizontal bar for each of the devices. First horizontal bar shows SMP task executions (original and creation tasks), last two bars show tasks running on shared locked resources (output DMA memory transfer from the FPGA, and DMA programming - submit) and the rest of the bars show `mxmBlock` tasks executed in the accelerators. The analysis shows that the current scheduling policy does not help to improve the performance when running `mxmBlock` in both SMP and FPGA. The high cost of executing the SMP version of the task compared to the FPGA version may be translated into a huge load imbalance problem if a wrong scheduler decision is taken. This has a significant impact in the case of 1 acc 128x128 running in both SMP and FPGA.

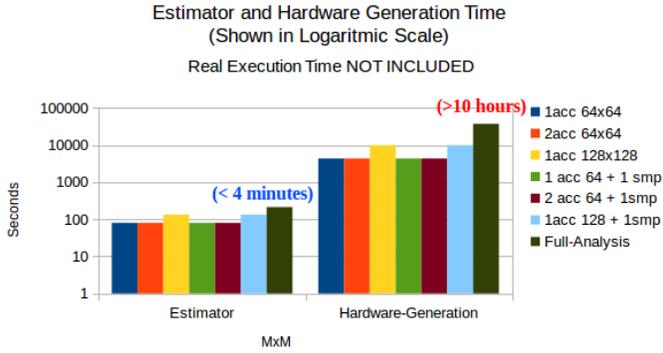

Fig. 6. Matrix Multiplication analysis time compared to hardware generation time of the hardware accelerators.

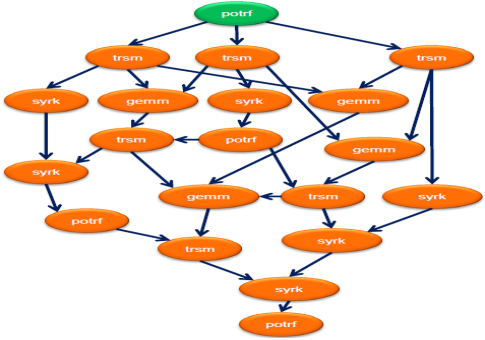

Fig. 8. Cholesky task dependecy graph for number of blocks equal to 4.

In the case of the tiled cholesky we have evaluated different resources distribution approaches between kernels that execute interleaved due to the complex nature of the cholesky dynamic graph (Figure 8). In order to simplify the example, the task granularity is fixed (64x64 blocks) and we have only evaluated which kernels should or should not be accelerated in the FPGA (note that to further complicate the scheduling, the application even has tasks - `dpotrf` task of Figure 4 - that can only be run in the SMP). As it is shown in Figure 9, the same speedup (normalized to the slowest configuration) trends are obtained in both estimated and real performance. The first three bars show the performance impact of implementing accelerators that try to maximize the usage of the hardware resources of the programmable logic (`FR-dgemm`, `FR-dsyrk`, `FR-dtrsm`, where `FR` stands for full resources), which limits the number of accelerators that fit in the hardware to one and forces all the other kernels to be executed in the SMP. The last set of three bars evaluate the performance of all the possible combinations of two tasks among three annotated with target FPGA (`dgemm+dgemm`, `dgemm+dsyrk`, `dgemm+dtrsm`) as the configuration only supports two accelerators.

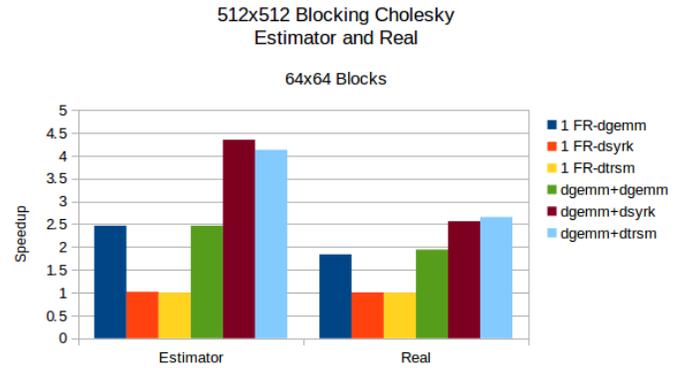

Fig. 9. Estimation and real cholesky performance comparison for different hardware configurations of the system and task configurations.

The programming productivity gain of the tiled cholesky is much more significant. A full analysis of those combinations requires one day and a half compared to less than 10 minutes with our methodology. Indeed, this day and a half is just for hardware generation time, no execution time is included neither creating the hardware design and integrating it to the rest of the application.

## VII. CONCLUSIONS AND FUTURE WORK

In this paper we have presented the current status of a heterogeneous parallel performance estimator that can help to potentially reduce the development effort in heterogeneous parallel computing systems like the Zynq All-Programmable SoC. The methodology is currently implemented for OmpSs applications and Zynq SoC. OmpSs is a task-based dataflow parallel programming model that helps to express heterogeneous task decompositions of an application. Thus, the programmer can annotate the application with OmpSs directives to identify tasks and the target devices where those tasks can be executed at run-time. Based on this information, our methodology estimates which is the best hardware-software partitioning of the annotated tasks on the Zynq Soc in few minutes. Results show that the best configurations and OmpSs annotations chosen by our estimation correspond with the real ones for the evaluated applications and configurations. And although the current performance estimator toolchain could be extended to automatically take care of different numbers of resources (e.g. the number of channels between the FPGA and the memory, cache coherence impact, etc), and explore different design space exploration strategies, the current implementation already shows speedups of more than two orders of magnitude (minutes vs days) on the process of achieving high heterogeneous performance for complex applications like cholesky.

Future work is to integrate power-efficiency and look-ahead scheduling heuristics into the simulator as well as helping the programmer with the hardware/software partitioning strategy to improve performance and/or area for a broader set of application domains.

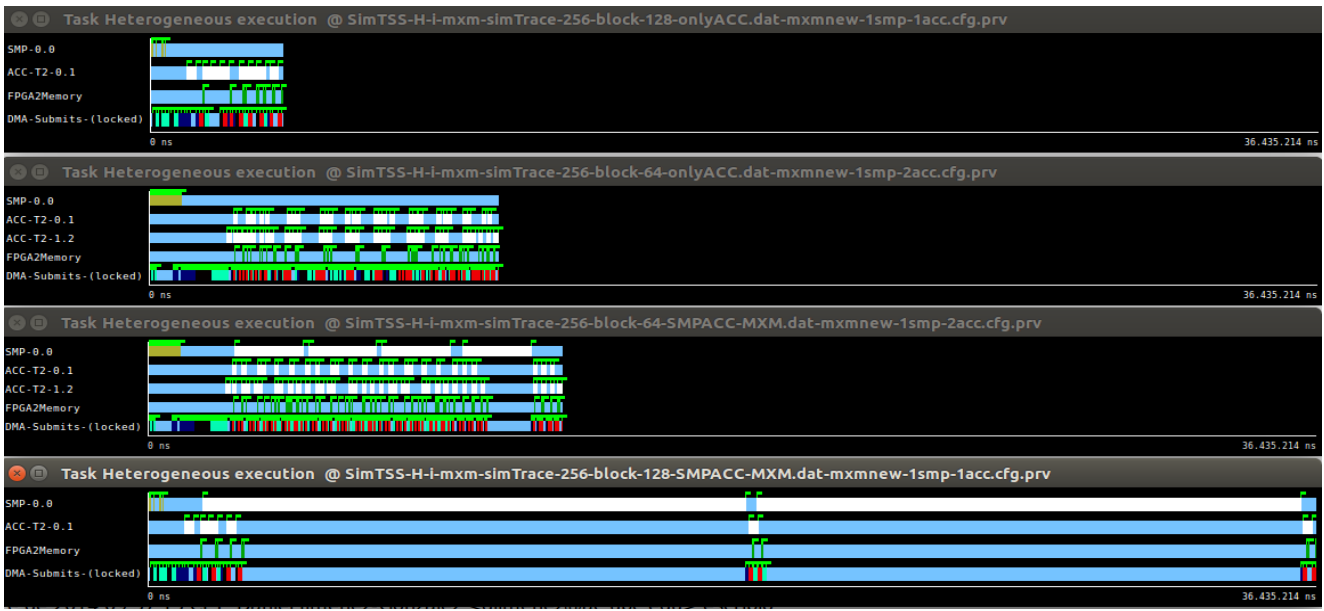

Fig. 7. MxM performance estimator traces for heterogeneous task executions running on 1 or 2 accelerators and none/one SMP. Blocksizes: 64x64 and 128x128.

## VIII. Acknowledgments

We thank the anonymous referees for their valuable feedback. This work is supported by the AXIOM project, funded by EU H2020 program (grant ICT-01-2014 GA 645496), the Spanish Government, through the Severo Ochoa program (grant SEV-2011-00067) the Spanish Ministry of Science and Technology (TIN2012-34557) and the Generalitat de Catalunya (MPEXPAR, 2014-SGR-1051). We thank the Xilinx University Program for its hardware and software donations.